\documentclass[conference]{IEEEtran}
\IEEEoverridecommandlockouts
\usepackage{cite}
\usepackage{amsmath,amssymb,amsfonts}
\usepackage{algorithm}
\usepackage{algpseudocode}
\usepackage{graphicx}
\usepackage{textcomp}
\usepackage{xcolor}
\usepackage{ntheorem}
\theoremseparator{:}

\usepackage{fnpct}
\ifCLASSOPTIONcompsoc \usepackage[caption=false,font=normalsize,labelfont=sf,textfont=sf]{subfig}
\else
\usepackage[caption=false,font=footnotesize]{subfig}
\fi
\usepackage{tikz}
\def\BibTeX{{\rm B\kern-.05em{\sc i\kern-.025em b}\kern-.08em
    T\kern-.1667em\lower.7ex\hbox{E}\kern-.125emX}}

\usepackage[nolist]{acronym}

\begin{acronym}
    \acro{ADN}{active distribution network}
    \acro{DER}{distributed energy resource}
    \acro{BGM}{balance group manager}
    \acro{BRP}{balance responsible party}
    \acro{ADMM}{alternating direction method of multipliers}
    \acro{TSO}{transmission system operator}
    \acro{LMP}{locational marginal pricing}
    \acro{TSO}{transmission system operator}
    \acro{SOE}{state-of-energy}
    \acro{GCP}{grid connection point}
\end{acronym}

\begin{document}

\title{Experimental Validation of Distributed Dispatching
of Multiple Active Distribution Networks Using the
ADMM\\
\thanks{This work was sponsored by the Swiss Federal Office of Energy’s “SWEET” programme and performed in the PATHFNDR consortium. The authors would also like to thank Severin Julen for his contribution in the early stages of the controller development.}
}

\author{\IEEEauthorblockN{1\textsuperscript{st} Matthieu Jacobs}
\IEEEauthorblockA{\textit{Distributed Electrical Systems Laboratory} \\
\textit{EPFL}\\
Lausanne, Switzerland \\
matthieu.jacobs@epfl.ch}
\and
\IEEEauthorblockN{2\textsuperscript{nd} Hanmin Cai}
\IEEEauthorblockA{\textit{Empa} \\
Dübendorf, Switzerland \\
hanmin.cai@empa.ch}
\and
\IEEEauthorblockN{3\textsuperscript{rd} Mario Paolone}
\IEEEauthorblockA{\textit{Distributed Electrical Systems Laboratory} \\
\textit{EPFL}\\
Lausanne, Switzerland \\
mario.paolone@epfl.ch}
}

\maketitle

\begin{abstract}
This paper presents the experimental validation of a framework for the coordinated dispatch and control of multiple \acp{ADN} hosting \ac{DER}. We show that the presented method, which builds further on work done in \cite{gupta_dd}, effectively allows to control multiple \acp{ADN} in a distributed way to ensure they achieve a common objective without revealing information on their DERs capabilities or grid model. This experimental validation is carried out using demonstrators at the DESL  
of EPFL and the NEST site at Empa, both in Switzerland. The coordination of the systems to share the flexibility of their controllable assets is demonstrated through a set of 24h experiments. Finally, the limitations of the method are discussed and future extensions proposed. 
\end{abstract}


\section{Introduction}
\acresetall
\label{sec:intro}
Research on distributed control applied to power systems has gained significant attention due to its application to various players including aggregators, prosumers, grid operators and more. 
The control of electric networks has been widely studied in literature. An important observation is that multiple players (e.g. aggregators, individual prosumers, system operators) are active in the electric power system and the control of the different assets thus naturally occurs in a distributed way. In what follows we specifically focus on the distributed control of assets in \acp{ADN}. This subject has been studied in literature, and \cite{distributed_opf} thoroughly reviewed existing OPF-based approaches. A natural application of distributed control in \acp{ADN} is the handling of distributed energy resources (DERs) such as EVs \cite{sherif_dev}. In both \cite{distributed_opf} and \cite{sherif_dev}, the authors consider multiple actors in one \ac{ADN}, with a central problem solved by the system operator. However, a realistic electricity systems contains multiple \acp{ADN} and, in general, they may be operated by different system operators, each optimizing their own system. In this respect, in another approach, all distribution system operators simply follow commands from a central coordinator \cite{tang_DERC}. It is, however, unclear how this coordinator obtains data on the capabilities and costs of the different \acp{ADN}. Multiple approaches to represent the aggregated flexibility capabilities of \acp{ADN} have been proposed. A central coordinator can then use these capabilities to schedule and control the ensemble of \acp{ADN}. For example the authors in \cite{nwcog} propose a method to advertise the maximum range of time-coupled active power realisations at the GCP of an \ac{ADN}, accounting for network constraints. However, this approach does not take into account the operational costs of the different flexibility assets, making the advertised capabilities cost-agnostic. In \cite{chen_flexagg} an economic condition is considered to advertise the flexibility generating the maximum benefit. This method, however, restricts the available flexibility area to a hyperbox. Alternatively, distributed optimization schemes can be used and the adoption of an \ac{ADMM}-type algorithm, to coordinate the usage of flexibility of different actors, allows to account for the cost of the available controllable resources, which is reflected through the dual variables, while accurately representing the network constraints. It should be noted that all of the works above use simulations to validate the developed methods and do not study the implementation of the proposed distributed control frameworks on physical systems. 

An interesting use case for distributed control of \acp{ADN}, is the concept of a balancing group employed in different market structures and recently extended to renewable energy communities \footnote{According to \cite{eu_dir}, these communities should be able to access all electricity markets and are financially responsible for imbalances they cause in the same way as a \ac{BRP}.}. This is a pool of consumers and producers represented by a \ac{BGM} or a \ac{BRP}. According to the definition in \cite{Swissgrid}, \acp{BRP} have to optimally balance between power demand and supply within their balance group at the time of delivery while ensuring compliance with schedule management requirements. The BRP holds the financial responsibility for imbalances settled by the national \ac{TSO} \cite{EC_balancing}. The different members of a balancing group may be geographically separated \cite{sg_balancing}. This use case requires controlling flexible \acp{DER} to adjust the aggregated power exchange of the balancing group. Through coordination between the different members of a balancing group, a solution that is technically feasible and economically optimal may be obtained. As the different \acp{ADN} in a balancing group do not disclose information about their assets, and thus prohibit the modeling of the full system by a single system operator, a coordination mechanism is required.

It is with a distributed structure in mind that the  methodology in \cite{gupta_dd} has been developed, where the authors propose to use an \ac{ADMM}-based approach to coordinate the scheduling of a MV system interfacing multiple LV \acp{ADN}. The goal of such an ADMM-based approach is to track an active power trajectory, called 
a dispatch plan, at the \ac{GCP} of the MV system. However, the approach required that each LV \ac{ADN} follows in the real time control stage their own energy exchange schedule defined day ahead. This reduces the coordination possibilities in real time between the different LV system operators or aggregators. Leveraging the possibility to exchange flexibility is precisely the goal of this work. Therefore, we remove the requirement that each \ac{ADN} must track its own share of the dispatch plan during the real time stage to fully exploit and mutualize the flexibilities of assets present in the different LV \acp{ADN}.

\par
Considering the identified gaps in the literature, the main contribution of this work is threefold:
\begin{enumerate}
    \item We address the limitation of the approach in \cite{gupta_dd} that requires fixed participant contributions in the day-ahead coordination.
    \item An intra-day tracking control algorithm is proposed, allowing participants to exchange flexibility in real time.
    \item The experimental validation of the proposed framework on a distributed set of real LV \acp{ADN}.
\end{enumerate}

\section{Problem Statement}
\label{sec:prob}
The problem tackled in this paper is the coordination of \acp{ADN} operated by different operators in a way that allows to track a common dispatch plan, computed as the aggregate power exchange at the grid connection point of the different considered \acp{ADN}. This problem is formulated in a distributed way using the \ac{ADMM}. Additionally, it is formulated in two stages, according to standard market practices. First, in the day-ahead scheduling stage, the different \acp{ADN} joining the control, coordinate to agree on a common dispatch plan to be announced to the balancing group for all the dispatching periods in the next day. The outcome of this first stage is a set of power exchange setpoints at each time step during the next day, which represents the net power exchanged by all the participants in this coordination framework. Next, in the intra-day stage, the different \acp{ADN} coordinate in real-time to determine how they will share the power agreed upon in the day-ahead stage. In this way, flexibility can be shared between the different \acp{ADN} and the controllable assets can be better used to track the agreed aggregated trajectory\footnote{As this approach does not account for the power flows on the upper-layer interconnecting grid, it may induce congestions. However, this is a second order problem typically solved by the upper-layer grid operator by redispatching or using LMPs.}.

\subsection{Use Cases}
% Different use cases are considered in which this approach can lead to increased flexibility of the \acp{ADN}. 
These two cases are studied to demonstrate the effectiveness of the proposed framework in increasing \acp{ADN} flexibility:
\begin{enumerate}
    \item Due to \acp{ADN} geographic separation, meteorological conditions may differ, leading to very different power profiles forecasted and considered in the day-ahead stage. Therefore, \acp{ADN} may coordinate their operation in real-time to minimise penalties for the balancing group due to mismatch from the scheduled dispatch plan. 
    \item The operating costs of the controllable assets in an \ac{ADN} may be significantly higher than in other \acp{ADN}. In this case, the flexibility of the controllable assets in \ac{ADN} hosting cheaper assets will be used preferentially to ensure reliable tracking of the aggregated dispatch plan.
\end{enumerate}

\subsection{Scheduling and Control Framework}
In this section, the day-ahead scheduling and the intra-day tracking frameworks are presented. In both problems, network constraints are modeled through iterative linearization of the power flow equations using the so-called sensitivity coefficients \cite{SensCoef}. Linear problems sequentially obtained admit a solution that is globally optimal for a given iteration, and through the iterative computation of the sensitivity coefficients, the power flow equations are inherently satisfied at convergence, although a global optimum at convergence cannot be guaranteed. In this work, only batteries are considered as controllable assets, while PV and loads are assumed to be inflexible and stochastic and lead to uncertainty. The battery losses are dealt with by considering a resistive line as in \cite{batterymodel}. The battery \ac{SOE} is constrained within the minimum and maximum allowable limits \footnote{The \ac{SOE} terminology is used as it is associated to the integral of the power exchanged by the battery with the network.}. Additionally, the reactive power injected by the battery inverter is related to the active power through a linear empirical relationship, such as presented, for example, in \cite{lin_approx}. This means that the constraints for each \ac{ADN} control problem are fully linear. In the following, we use the subscript $n$ to refer to the concerned \ac{ADN} and $x_n$ to gather the variables for \ac{ADN} $n$. 

\subsubsection{Scheduling Stage}
The scheduling stage is formulated as a stochastic optimization problem, meaning the vector variables $\boldsymbol{x_n}$ contain the variables representing the controllable injections, state variables and respective \ac{ADN}'s slack power in the set of scenarios $\Omega$. A scenario-based approach is used to sample generic probability density functions of stochastic resources. More specifically, by using global horizontal irradiance (GHI) forecasts from the German weather service DWD (detailed in \cite{dwd}), consistent profiles for the different locations are obtained. The GHI predictions are provided as a set of scenarios generated through ensemble forecasts obtained from a set of different simulation models. These can directly be used as inputs for the stochastic optimization in the day-ahead problem.  For the loads,  20 scenarios are considered in the day-ahead stage based on historical residential load profiles. In the scheduling stage, the goal is to obtain the aggregated dispatch plan of the concerned \acp{ADN} $P_d = \sum_n x^s_n$. 
To obtain the optimal solution for the group of collaborating \acp{ADN}, the \ac{ADMM} algorithm is used. As known, the ADMM involves the presence of a main and sub-problems. The global power balance is defined in the main problem, relating the power exchange of each \ac{ADN} with the upper-level network to the total power exchanged by the group of considered \acp{ADN}, denoted as $P_s$. The variables representing the power exchange of each  \ac{ADN} are shared with the main problem and referred to as $x^{s}_n$, while the subproblem private variables are referred to as $x^p_n$. The shared variables $x^{s}_n$ are duplicated to obtain the copied variables for the ADMM algorithm. The duplicates are represented by $\hat{x}^s_n$. These variables allow to share information between the subproblems and the main problem of the \ac{ADMM} algorithm. Additionally, we define the auxiliary variables $r$ representing the ramping of the aggregated dispatch plan and specify the variables $p^b_n$ as a part of the private variables of each  \ac{ADN}, representing the controllable injections. A coefficient $W$ is used in the objective to increase the weight of the dispatch tracking objective. The main problem can then be written as follows with $u_n$ being the auxiliary parameter introduced by the ADMM algorithm.

\begin{align}
    &\min_{\boldsymbol{\hat{x}^{s}}, P_d}  W ||P_s - P_d||^2 + ||r||^2 +\frac{\rho}{2} \sum_n(||x^{s}_n - \hat{x}^{s}_n + u_n||^2)\\
   &  \textrm{s.t.} \quad  \sum_n(\hat{x^{s}_n}(t,s))  + P_s(t,s) = 0 \quad \forall t \in \mathcal{T},\forall s \in \Omega \\
    &r(t,s) \geq P_s(t,s)-P_s(t-1,s) \quad \forall t \in \mathcal{T},\forall s \in \Omega\\
    & r(t,s) \geq -(P_s(t,s)-P_s(t-1,s)) \quad \forall t \in \mathcal{T},\forall s \in \Omega
\end{align}

With the introduction of coefficients $w_b$ to weigh the operational costs of the different batteries, the subproblems are written in (\ref{eq:sub}) with the linear grid model and resource constraints collected in (\ref{eq:sub_con}).

\begin{align}
\label{eq:sub}
    \min_{x^s_n} & \frac{\rho}{2} * ||x^{s}_n - \hat{x}^{s}_n + u_n||^2 + ||r^{s}_n||^2 + \sum_b w_b ||p^{b}_n||^2 \\
    \label{eq:sub_con}
    & \textrm{s.t.} \quad  A_n * x_n \leq b_n
\end{align}

The main problem and the different subproblems are then solved in an iterative way, as described in Algorithm \ref{alg:ADMM}. For simplicity, we introduce a variable $x^s$ that includes the shared variables $x^s_n$ for all considered subproblems. The algorithm defines the primal and dual residuals $r$ and $s$, which are updated every iteration to quantify the distance from convergence.

\begin{algorithm}[h]
\caption{ADMM algorithm of the day-ahead scheduling}
\label{alg:ADMM}
\begin{algorithmic}[1]
\Require {GHI\_forecasts, load\_forecasts $\forall$ ADNs}
   \While{k $\leq$ k\_max and not converged}
    \State $\boldsymbol{\hat{x}^{s^k}}$ $\leftarrow$  main problem \;
    \State $\boldsymbol{x^{s^k}}$ $\leftarrow$ subproblems \;
    \State $ \boldsymbol{u^{k}} \leftarrow  \boldsymbol{u^{k-1}} + \boldsymbol{x^{s^k}} - \boldsymbol{\hat{x}^{s^k}}$ \;
    \State $r^k \leftarrow ||x^{s^k} - \hat{x}^{s^k}||$ \;
    \State $s^k \leftarrow \rho*||\hat{x}^{s^k} - \hat{x}^{s^{k-1}}||$ \;
    \State $r^k_{\mathrm{max}} \leftarrow \epsilon^{\mathrm{abs}} * n + \epsilon^{\mathrm{rel}} * $max$(||\boldsymbol{\hat{x}^{s^k}}||, ||\boldsymbol{x^{s^k}}||)$
    \State $s^k_{\mathrm{max}} \leftarrow \epsilon^{\mathrm{abs}} * n + \epsilon^{\mathrm{rel}} * $max$(||\boldsymbol{u^k}||)$
    \If{$r \leq r^k_{\mathrm{max}}$  and $s \leq s^k_{\mathrm{max}}$} \;
    converged $\leftarrow$ True
    \EndIf
    \EndWhile
    \end{algorithmic}
\end{algorithm}

\subsubsection{Tracking Stage}
In the tracking stage, the goal is to track the dispatch plan obtained in the scheduling stage. This is achieved through a bilevel MPC formulation. The stochastic variables are represented by their expectation and the receding horizon of the MPC accounts for the uncertainty. The contributions of each  \ac{ADN} are once again obtained by splitting the problem in a main one and a set of subproblems through an \ac{ADMM} approach that is formulated with a shrinking horizon until the end of the day. The results of the MPC problem are the setpoints for the controllable variables of each  \ac{ADN}, from which the estimated slack power for each  \ac{ADN} follows. Within each dispatching period, the dispatching error is recomputed (for instance every minute) and the \acp{ADN} solve the lower-layer \ac{ADMM} MPC to determine the setpoint adjustments for the next minute. This ensures the effect of prediction errors to be minimized and the aggregate energy exchanged within a dispatching period to be close to the agreed dispatch plan. It is worth noting that the problems solved in the tracking stage must be very similar to the ones solved in the scheduling stage to guarantee consistency between the day-ahead and the tracking stage. The constraints are the same, with the only difference that they are formulated including the expectation of the stochastic variables. The objectives are, however, slightly different, as the dispatch plan is now a constant and no longer the decision variable. The objective for the main problem then becomes:

\begin{align}
    \min_{\boldsymbol{\hat{x}^{s}}_n} & W ||P_s - P_d||^2 + ||r||^2 + \frac{\rho}{2} \sum_n(||x^{s}_n - \hat{x}^{s}_n + u_n||^2)
\end{align}

The subproblem objectives then have the following form:
\begin{align}
    \min_{x^{s}_n} & \frac{\rho}{2} ||x^{s}_n - \hat{x}^{s}_n + u_n||^2 + \sum_b w_b ||p^{b}_n||^2 
\end{align}

The upper- and lower-layer MPC problems solved during the tracking stage show a subtle difference. In the upper layer, the variable $x^s_n$ represents the slack power of the $n$th \ac{ADN} , while in the lower layer it represents the adjustment of the slack power within the dispatching period to compensate for the energy mismatch incurred during the current period. The minimization of these adjustments is captured in the objective of the main problem. The solution of the bilevel MPC problem gives the setpoints for the controllable assets in each  \ac{ADN}.

\section{Experimental Setup}
\begin{figure}[h]
    \centering
    \includegraphics[width=0.65\linewidth]{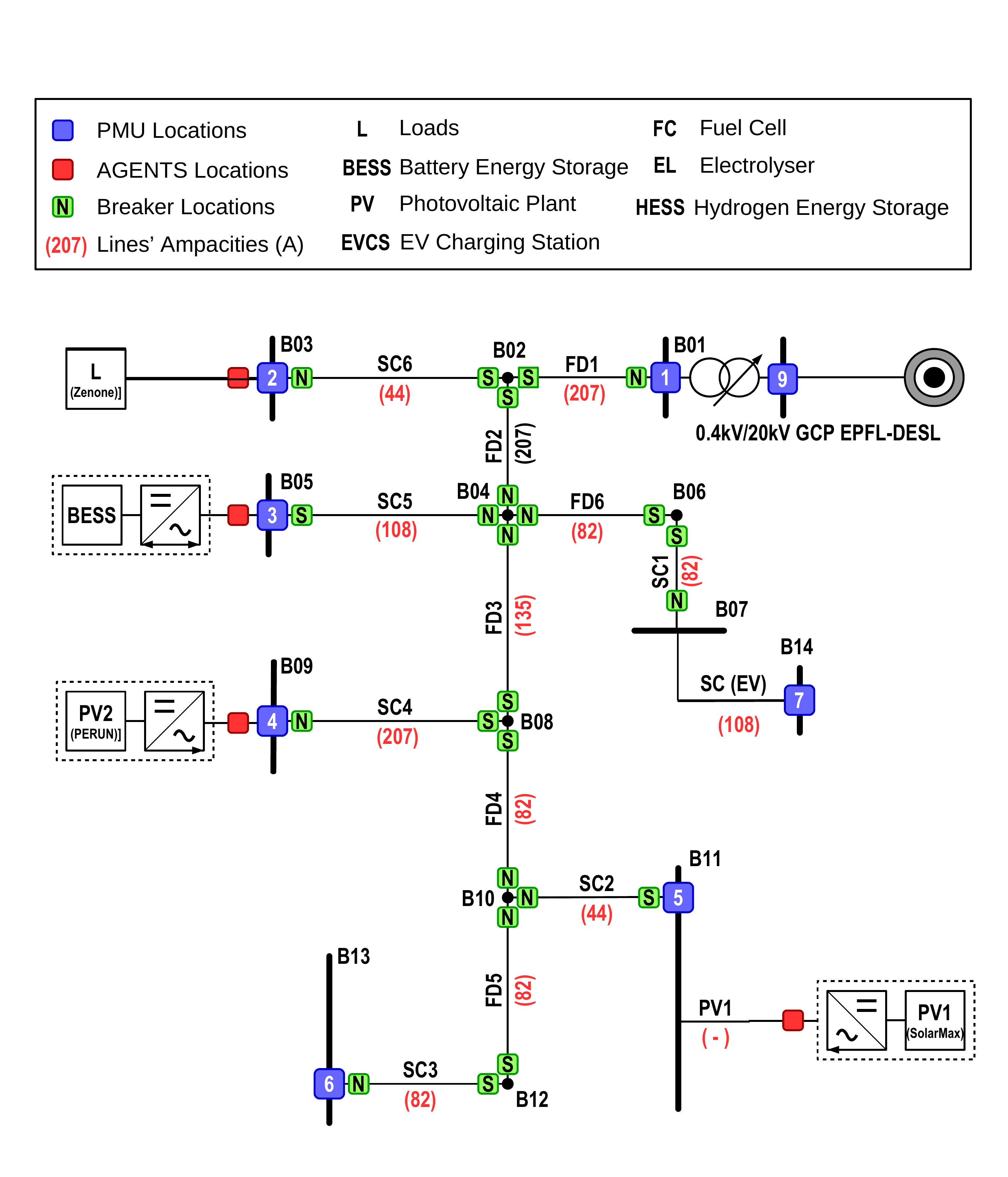}
    \caption{Configuration of the EPFL-DESL LV microgrid.}
    \label{fig:DESL}
\end{figure}

\begin{figure}[h]
    \centering
    \includegraphics[width=0.6\linewidth]{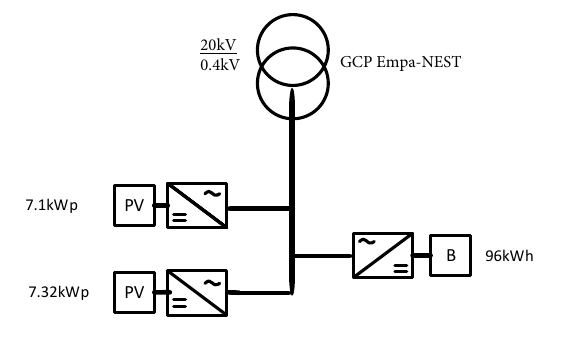}
    \caption{Configuration of the Empa-NEST LV microgrid.}
    \label{fig:EMPA}
\end{figure}

We consider two independent systems, in this case LV \acp{ADN}, which are assumed to be part of the same balancing group. The DESL microgrid at EPFL, which is shown in Figure \ref{fig:DESL} and a subset of the system of Empa-NEST building ~\cite{richner2018nest}, as shown in Figure \ref{fig:EMPA}. Both systems contain battery energy storage system (BESS) and PV panels with a significant amount of stochastic electricity production. Additionally, in the EPFL-DESL microgrid an uncontrollable residential load is considered. The PV production is considered to be uncontrollable, as the goal in this work is to focus on the sharing of limited sources of flexibility in a coordinated way to allow optimal grid operation. 

\subsection{Information Flow}
\begin{figure}[h]
    \centering
    \includegraphics[width=0.95\linewidth]{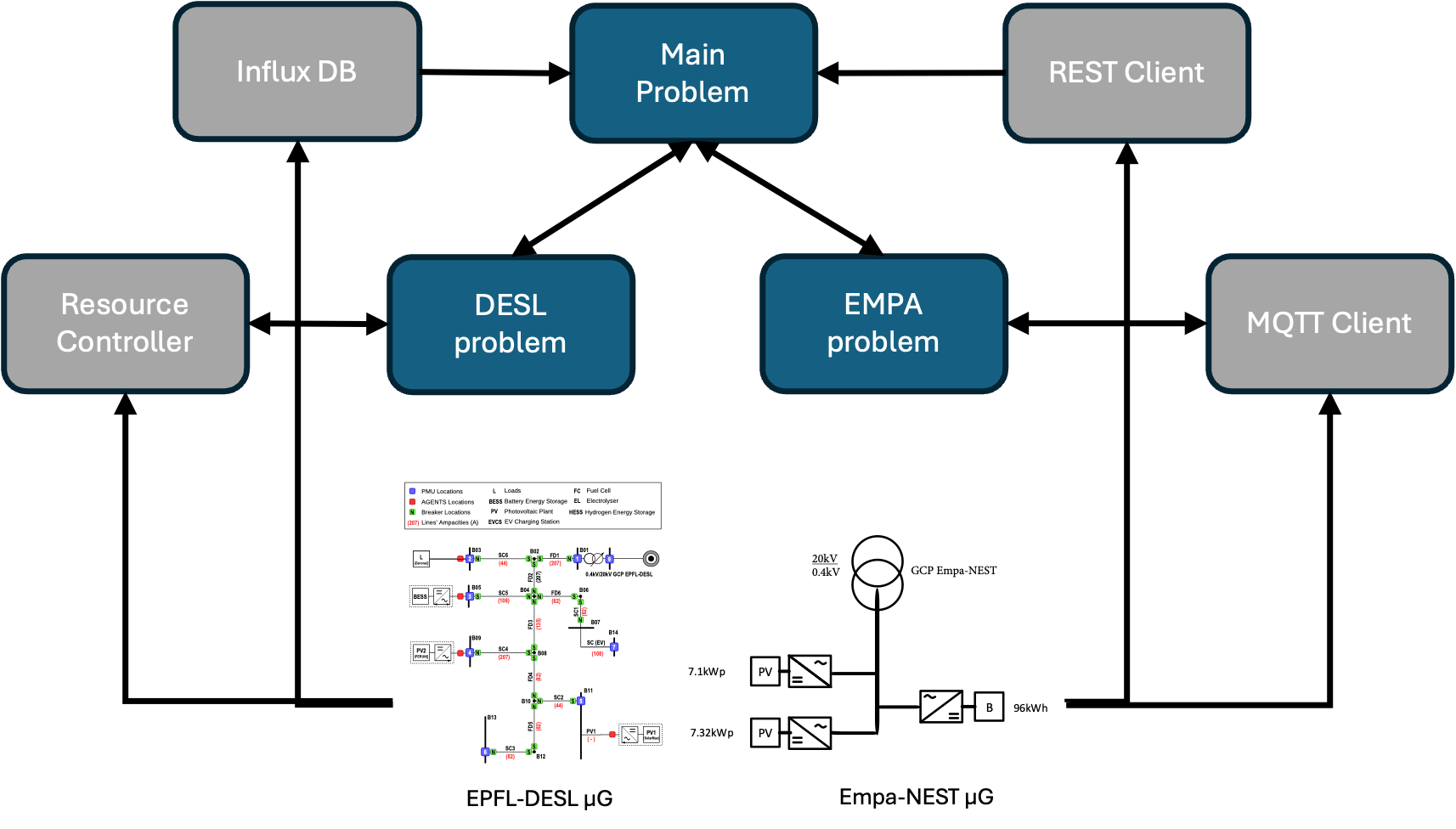}
    \caption{Communication architecture with information flows between the main problem and the sub problem.}
    \label{fig:comms}
\end{figure}

A schematic representation of the information flow is provided in Figure \ref{fig:comms}. Resource and network state measurements are obtained from the resource controller or the message queuing telemetry transport (MQTT) client, on the EPFL-DESL and Empa-NEST side, respectively, while aggregate values for the slack power of  each \acp{ADN} in the current dispatching period are obtained through the Influx DB database or the representational state transfer (REST) client. The main problem and the subproblems for both networks are solved iteratively until convergence or the iteration limit and then the setpoints from the subproblems are sent back to the resource controller/MQTT client for actuation. 
\subsection{Experiment}
The presented experiments are carried out over a period of 24 hours starting at midnight. The dispatch plan is computed once and the computation starts at 23:30. The SOE for the batteries is initialized in the center of their operating ranges.
The dispatch plan has a resolution of 15 minutes, according to standard market rules and it is the average power exchange within such a period that is of interest for the evaluation of the intra-day tracking. 
Considering the stochastic generation through PV, predictions from the DWD are considered. The stochastic load is implemented through a load emulator where an out-of-sample historical realization is then applied for the experimental validation of the tracking scheme.

\section{Results and Discussion}
\label{sec:results}
\subsection{Dispatch Tracking}
\begin{figure}[h]
    \centering
    \includegraphics[width=0.85\linewidth]{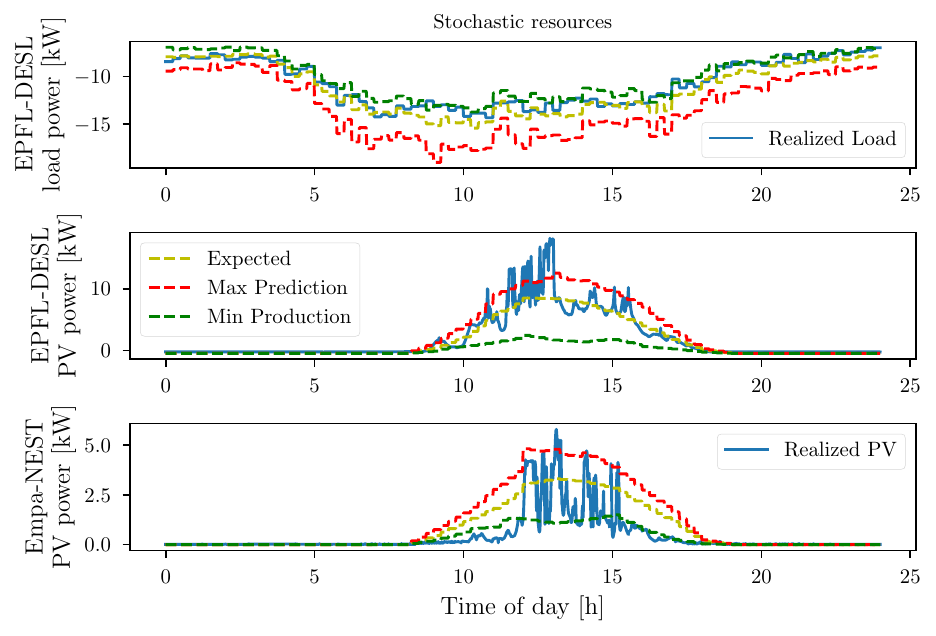}
    \caption{Prosumption profiles of stochastic \acp{DER} for the coordination of similar \ac{ADN}s, shown at one minute resolution. The top plot shows the load of the EPFL-DESL microgrid, defined as a negative injection. The next two show the total PV injection at the EPFL-DESL and Empa-NEST microgrids respectively.}
    \label{fig:stochastic}
\end{figure}

\begin{figure}[h]
    \centering
    \includegraphics[width=0.85\linewidth]{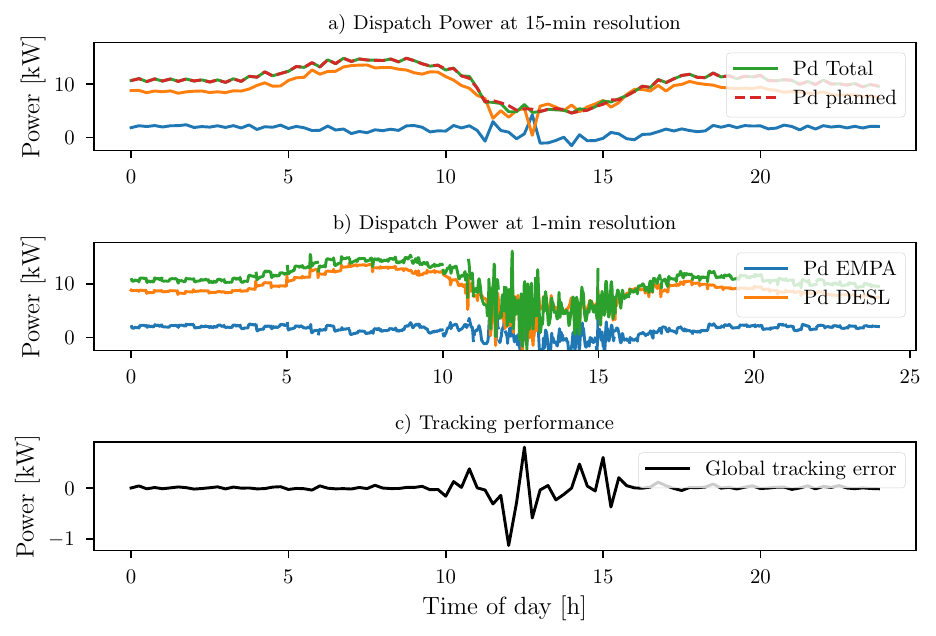}
    \caption{Dispatch tracking performance over 24 hours for the coordination case. The power exchanged by both the EPFL-DESL and the Empa-NEST systems are also shown.}
    \label{fig:dispatching}
\end{figure}

The realized PV and load profiles are presented in Figure \ref{fig:stochastic}, together with the minimum and maximum values from the considered scenarios at each time step (represented by dashed lines).
Figure \ref{fig:dispatching} shows the realized aggregated power exchange with the upper-layer network. The values shown are the fifteen minute averages and they are to be compared with the dispatch plan agreed upon by the different \acp{ADN} in the day-ahead stage. The contribution of each \ac{ADN} is also shown. 

\begin{figure}[h]
    \centering
    \includegraphics[width=0.85\linewidth]{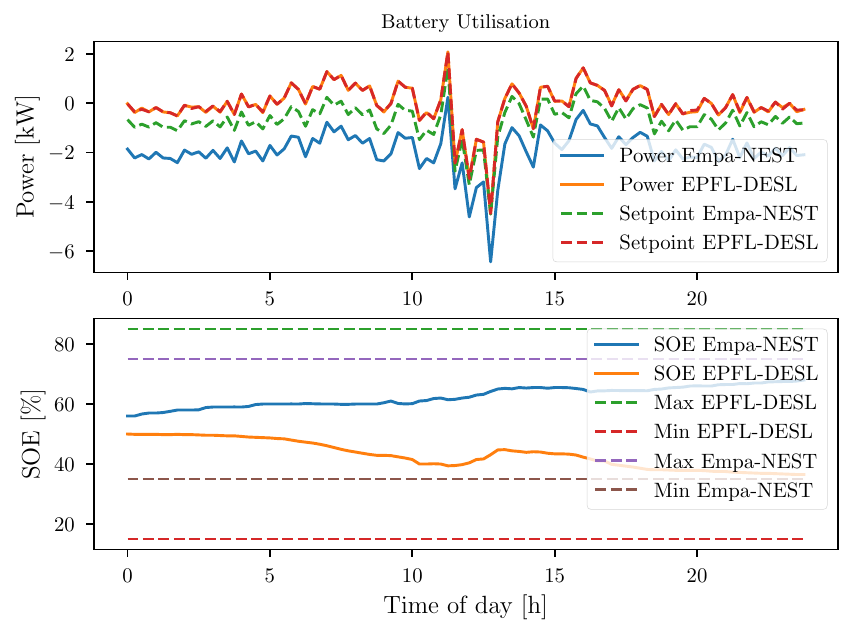}
    \caption{Controllable resources actuation. The top figure shows the power injections for the batteries in the EPFL-DESL and Empa-NEST microgrids respectively. The bottom plot shows the SOE for these batteries.}
    \label{fig:actuation}
\end{figure}

In Figure \ref{fig:actuation}, we also show the power injections of the controllable assets. This shows that, although both batteries follow the same trajectory, there is an offset between the power injections. This is due to a the actuation precision of the Empa-NEST battery as there is an offset between the requested setpoint and the realised power injection. The results clearly show that the approach allows to accurately track the dispatch plan computed in the day-ahead stage. The actuation of the batteries also demonstrates that the algorithm allows to effectively coordinate the operation of the two systems as both batteries are used to compensate for prosumption fluctuations, even when these are present in only one of the systems.

\subsection{Tracking error reduction and the impact of delays}
In the tracking stage, the goal is to minimize the error between the 15 minute average dispatch power and the dispatch plan. To this end, the lower layer MPC takes the energy error incurred in the present dispatching period and attempts to compensate for this error in the remaining time steps of that period. Compensating the full error in the next time step would lead to large oscillations due to the fast variations of the PV injections and the delays in the measurements. Therefore, the error compensation is spread out over the remaining time steps of the dispatching period and adjusted during every lower layer MPC computation. A linearly decaying coefficient was found to yield the best behavior of the real time control layer.
\newline Tracking errors may occur due to prediction errors and delays. Prediction errors are inherently present due to the stochastic character of the prosumption. The delays impacting the tracking are twofold. The first is due to the computation of the \ac{ADMM} algorithm. As this is an iterative algorithm, it requires a certain amount of time to converge to a solution on which all the members agree. The second source of delay is the measurement update. This delay is comparatively more pronounced for the system at Empa-NEST, as the measurements has a one-minute resolution. In other words, measurement within the minute is only available in the next minute and the action will be delayed by at least one full minute. To alleviate the effects of this delay, average values over the last five minutes are used for the PV production to reduce oscillatory behavior due to fast power variations. Additional delays in the actuation of the set point and the communication between the main optimizer and the resource controllers are also present, but these are negligible in comparison with the two mentioned above. A third source of tracking errors is due to the actuation precision of the battery at Empa-NEST as mentioned above.

\subsection{Flexibility Sharing}
\begin{figure}[h!]
    \centering
    \includegraphics[width=0.85\linewidth]{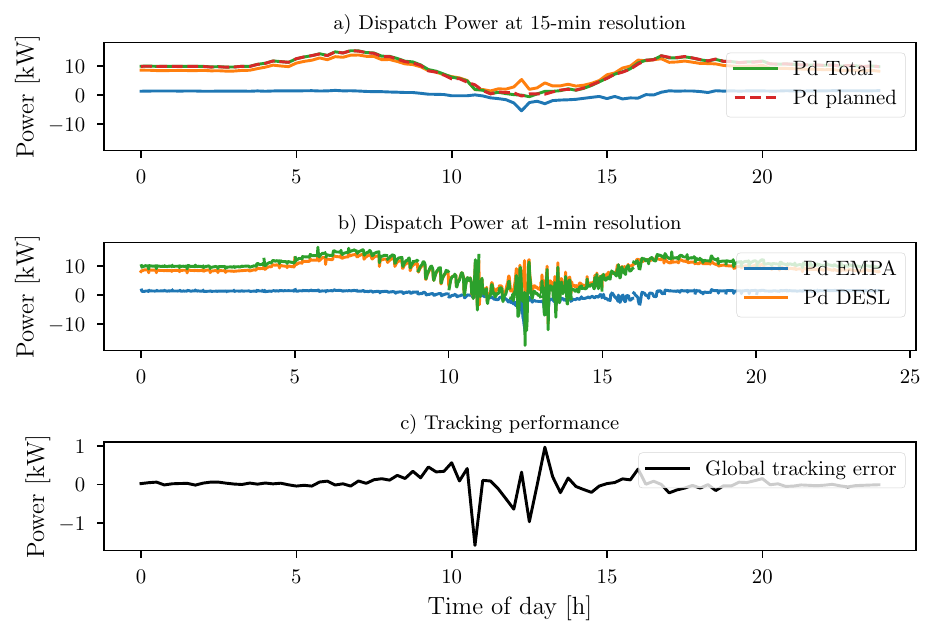}
    \caption{Dispatch tracking for case with flexibility price differences for the controllable assets.}
    \label{fig:mutuation_dispatch}
\end{figure}

\begin{figure}[h!]
    \centering
    \includegraphics[width=0.85\linewidth]{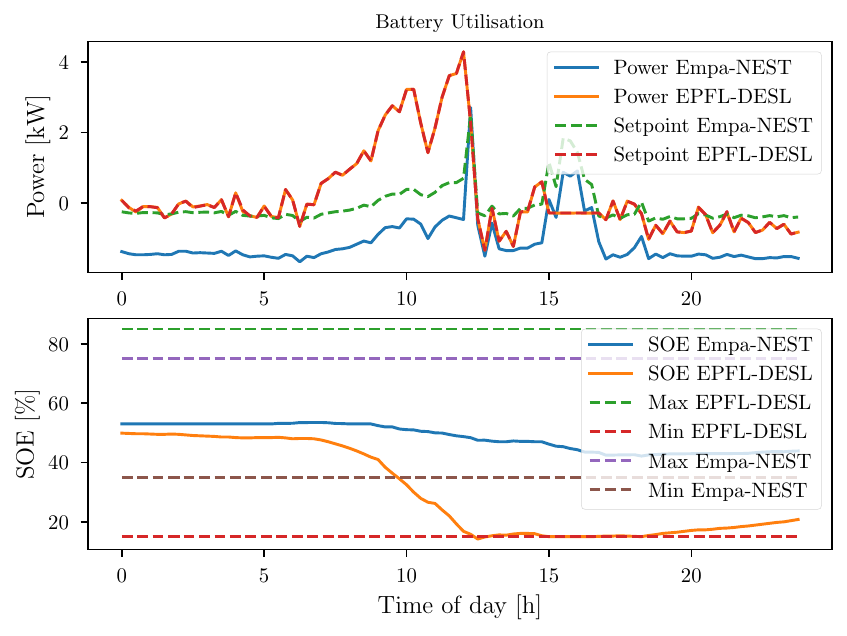}
    \caption{Actuation of the controllable assets with different operating costs.}
    \label{fig:mutuation_actuation}
\end{figure}

\begin{figure}[h!]
    \centering
    \includegraphics[width=0.85\linewidth]{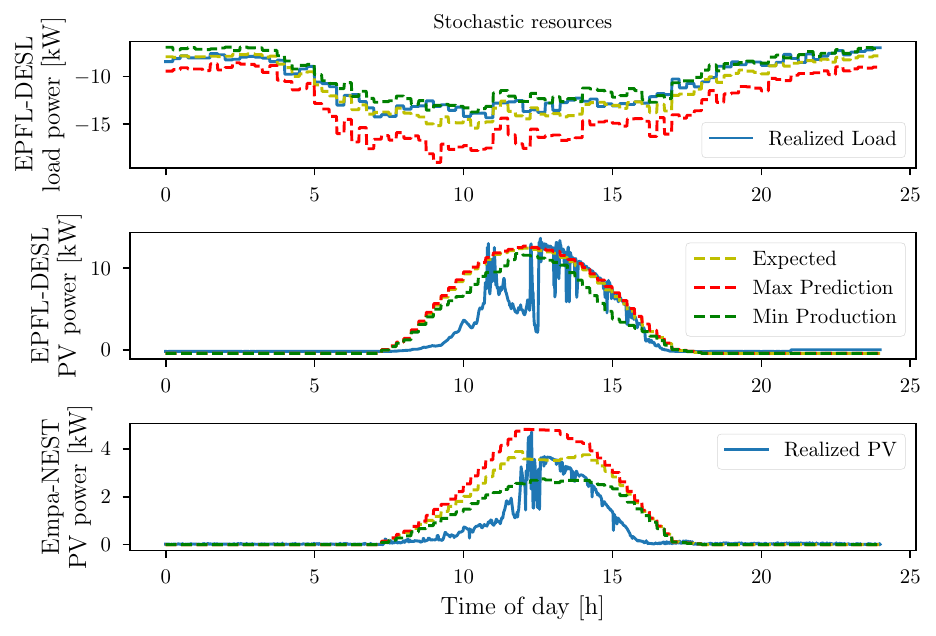}
    \caption{Prosumption profiles for the case study with different flexibility costs, shown at one minute resolution.}
    \label{fig:stochastic_bis}
\end{figure}

The presented experiment shows the coordinated approach using the \ac{ADMM} algorithm allows to effectively share the flexibility of the controllable resources present in the different \acp{ADN}. To emphasise this, we show results when the costs for providing flexibility are not the same for the assets in the different \acp{ADN}. This can be studied by varying the coefficient $w_b$ in the subproblem objective. Figure \ref{fig:mutuation_actuation} shows that cheap flexibility is used preferentially and more expensive flexibility only when required. The setpoint for the more expensive battery at Empa-NEST is zero until midday, when it starts being used due to saturation of the \ac{SOE} of the EPFL-DESL battery. This shows that the flexibility is indeed shared between the different participants of the balancing group. It is also clear that the cheaper battery (from EPFL-DESL) is used in a more dynamic way, showing the benefits of coordinated dispatch tracking. Additionally, Figure \ref{fig:mutuation_dispatch} shows that the aggregated slack power still tracks the dispatch plan. It is also clear that due to significant under performance of the PV generation during the first part of the day, this battery quickly discharges. Due to this, from the middle of the day onward, the more expensive battery is also actuated to ensure the dispatch plan is tracked. Figure \ref{fig:mutuation_dispatch} shows the (aggregated) energy exchanged by the balancing group at a minute resolution. Clearly, due to the inexactness of the forecast and the delays in actuation, the instantaneous power exchange will not always correspond to the dispatch plan but reflect the volatility of the prosumption. However, as shown above, the 15 minute averages, which are relevant when computing penalties, do track the scheduled dispatch. Finally, the stochastic injections for this second experiment are presented in Figure \ref{fig:stochastic_bis}. This shows an overestimation of the PV production on both sides in the morning and high volatility from 12pm.

\section{Conclusions and Future Work}
\acresetall
The experimental campaign presented in this paper demonstrate the effectiveness of the \ac{ADMM} in coordinated dispatching of multiple \acp{ADN} within a balancing group. The aggregated dispatch plan is tracked under conditions of high prosumption volatility and the coordination between multiple \acp{ADN} allows to utilize the cheapest source of flexibility to manage the uncertainties of all systems. Small errors may still occur when large fluctuations are present at the end of dispatching periods. This is due to the limited speed of the iterative control algorithm and the delay in acquiring the latest measurements. A faster data acquisition and exchange system would allow to change the setpoint more regularly and have measurements that are closer to the actual state. This will directly influence the tracking error as this is a function of the magnitude of the slack power mismatch and the time duration of this mismatch. 
Future work will include the demonstration of other aggregation schemes such as representing active distribution systems as virtual power plants to control a set of distributed assets to provide grid services such as primary frequency regulation superposed to the aggregated dispatching. 
\label{sec:conclusion}

\bibliographystyle{IEEEtran}
\bibliography{main.bib}

% Generated by IEEEtran.bst, version: 1.14 (2015/08/26)
\begin{thebibliography}{10}
\providecommand{\url}[1]{#1}
\csname url@samestyle\endcsname
\providecommand{\newblock}{\relax}
\providecommand{\bibinfo}[2]{#2}
\providecommand{\BIBentrySTDinterwordspacing}{\spaceskip=0pt\relax}
\providecommand{\BIBentryALTinterwordstretchfactor}{4}
\providecommand{\BIBentryALTinterwordspacing}{\spaceskip=\fontdimen2\font plus
\BIBentryALTinterwordstretchfactor\fontdimen3\font minus \fontdimen4\font\relax}
\providecommand{\BIBforeignlanguage}[2]{{%
\expandafter\ifx\csname l@#1\endcsname\relax
\typeout{** WARNING: IEEEtran.bst: No hyphenation pattern has been}%
\typeout{** loaded for the language `#1'. Using the pattern for}%
\typeout{** the default language instead.}%
\else
\language=\csname l@#1\endcsname
\fi
#2}}
\providecommand{\BIBdecl}{\relax}
\BIBdecl

\bibitem{gupta_dd}
R.~Gupta, S.~Fahmy, and M.~Paolone, ``Coordinated day-ahead dispatch of multiple power distribution grids hosting stochastic resources: An admm-based framework,'' \emph{EPSR}, vol. 212, p. 108555, 2022.

\bibitem{distributed_opf}
E.~Dall'Anese, H.~Zhu, and G.~B. Giannakis, ``Distributed optimal power flow for smart microgrids,'' \emph{IEEE TSG}, vol.~4, no.~3, 2013.

\bibitem{sherif_dev}
S.~Fahmy, R.~Gupta, and M.~Paolone, ``Grid-aware distributed control of ev charging stations in active distribution grids,'' \emph{EPSR}, 2020.

\bibitem{tang_DERC}
Z.~Tang, T.~Liu, C.~Zhang, Y.~Zheng, and D.~J. Hill, ``Distributed control of active distribution networks for frequency support,'' in \emph{PSCC}, 2018.

\bibitem{nwcog}
B.~Cui, A.~Zamzam, and A.~Bernstein, ``Network-cognizant time-coupled aggregate flexibility of distribution systems under uncertainties,'' in \emph{2021 American Control Conference (ACC)}, 2021, pp. 4178--4183.

\bibitem{chen_flexagg}
X.~Chen, E.~Dall’Anese, C.~Zhao, and N.~Li, ``Aggregate power flexibility in unbalanced distribution systems,'' \emph{IEEE TSG}, vol.~11, no.~1, 2020.

\bibitem{eu_dir}
E.~Parliament, ``Eu parliament, directive (eu) 2019/944 on common rules for the internal market for electricity.'' EU Parliament, Tech. Rep., 2019.

\bibitem{Swissgrid}
Swissgrid, ``Industry recommendation for the swiss power market balancing concept switzerland principles of balance management of the swiss power market,'' 2019.

\bibitem{EC_balancing}
C.~I.~R. (EU), ``Commission regulation (eu) 2017/2195 of 23 november 2017 establishing a guideline on electricity balancing,'' 2021.

\bibitem{sg_balancing}
Swissgrid, ``General balance group regulations (v2.5),'' 2024.

\bibitem{SensCoef}
K.~Christakou and e.~al, ``Efficient computation of sensitivity coefficients of node voltages and line currents in unbalanced radial electrical distribution networks,'' \emph{IEEE TSG}, vol.~4, no.~2, 2013.

\bibitem{batterymodel}
E.~Stai, L.~Reyes-Chamorro, F.~Sossan, J.-Y. Le~Boudec, and M.~Paolone, ``Dispatching stochastic heterogeneous resources accounting for grid and battery losses,'' \emph{IEEE TSG}, vol.~9, no.~6, pp. 6522--6539, 2018.

\bibitem{lin_approx}
M.~Nick, R.~Cherkaoui, and M.~Paolone, ``Optimal allocation of dispersed energy storage systems in active distribution networks for energy balance and grid support,'' \emph{IEEE TPS}, vol.~29, no.~5, 2014.

\bibitem{dwd}
A.~Paxian and B.~M. et~al., ``The dwd climate predictions website: Towards a seamless outlook based on subseasonal, seasonal and decadal predictions,'' \emph{Climate Services}, 2023.

\bibitem{richner2018nest}
P.~Richner, P.~Heer, R.~Largo, E.~Marchesi, and M.~Zimmermann, ``Nest--una plataforma para acelerar la innovaci{\'o}n en edificios,'' \emph{Informes de la Construcci{\'o}n}, vol.~69, no. 548, p. 222, 2018.

\end{thebibliography}

\end{document}